\documentclass[aps,prl,preprint,superscriptaddress]{revtex4-1}
\usepackage{graphicx}
\usepackage{color}
\usepackage[utf8]{inputenc}

\begin{document}

\title{Tunable photonic crystals explained via Mie theory and discrete dipole approximation: a different light}

\author{Luis J. Mendoza Herrera}%
\email{joaquinm@ciop.unlp.edu.ar}
\affiliation{Centro de Investigaciones Ópticas (CIOp), (CONICET La Plata-CIC), Argentina}
\affiliation{Departamento de Ciencias Básicas, Facultad de Ingeniería, UNLP, Argentina}

\author{Ignacio J. Bruvera}
\affiliation{Instituto de Física La Plata (IFLP), UNLP-CONICET}
\affiliation{Departamento de Física, Facultad de Ciencias Exactas, UNLP, Argentina}

\author{Lucía B. Scaffardi}
\affiliation{Departamento de Ciencias Básicas, Facultad de Ingeniería, UNLP, Argentina}

\author{Daniel C. Schinca}
\email{daniels@ciop.unlp.edu.ar}
\affiliation{Departamento de Ciencias Básicas, Facultad de Ingeniería, UNLP, Argentina}

\date{\today}

\begin{abstract}
A novel hybrid method based on Mie theory and the Discrete Dipole Approximation (DDA) was developed to study the microscopic parameters governing the optical response of tunable photonic crystals (PC). The method is based on a two-step process. An effective polarizability derived from Mie theory is determined by equating the extinction efficiency of an isolated nanoparticle (NP) to the extinction efficiency of an equivalent particle considering the dipolar limit. Then, this effective polarizability is used in the DDA framework to compute the optical response of an interacting particle array constituting the PC structure.   
	
As a particular example, the method was applied to a linear array of core-shell magnetite@silica NPs to study the dependence of extinction and absorption on system parameters such as core radius, shell thickness, total radius, interparticle separation, and size distribution. The results indicate that an increase in these parameters leads to a redshift of the extinction peak as well as an increase in its $FWHM$. Finally, the method is applied to fitting experimental results on reflection/transmission measurements of magnetite@silica NPs colloids subjected to different magnetic field strengths with very good agreement.
		
The presented method reduces the computational cost and time for the NPs sizes considered, and can be applied to PCs responsive to different stimuli such as mechanical stress, electric field and temperature, \textit{inter alia}.

\end{abstract}


\maketitle

\section{Introduction}
\section{\label{intro}}


Photonics crystals (PCs) represent the optical equivalent of electronic semiconductors, characterized by a spatially periodic dielectric function that modifies light propagation. These materials exhibit photonic band gaps, which prevent the transmission of specific frequencies in certain directions, enabling unprecedented control over electromagnetic waves \cite{RefB1}. The concept of PCs was first proposed independently by Yablonovitch and John in 1987, who theorized that periodic dielectric structures could create band gaps for photons, analogous to electronic band gaps in semiconductors \cite{RefY1}. Since their inception, PCs have revolutionized photonics, finding applications in low-loss optical fibers, high-quality resonant cavities, optical sensors, and light-emitting devices such as LEDs and lasers \cite{wang2023design, lonergan2023many, chaudhary2022advances}.

Recent advances by Tran \textit{et al.} \cite{RefT2} demonstrated the rapid assembly of magnetoplasmonic photonic arrays capable of producing bright, non-iridescent structural colors with responsiveness to external stimuli. These systems mark a significant breakthrough in photonic crystals as they enable dynamic control of optical properties via external magnetic fields. As highlighted by Yang et al. \cite{RefT3}, the use of anisotropic building blocks for the assembly of these materials has opened new avenues for the design of advanced functionalities photonic systems.

A key advantage of PCs lies in their ability to confine and guide light at subwavelength scales, making them indispensable for integrated photonics and quantum optics \cite{RefT4}. Advances in nanofabrication techniques, including electron-beam lithography and self-assembly, have enabled the realization of 1D, 2D, and 3D PC structures, each with unique optical properties \cite{cai2021colloidal, dai2020nanoscale}. Recent developments have further expanded their functionality through stimuli-responsive PCs, which dynamically alter their optical properties in response to external factors such as temperature, mechanical stress, and magnetic fields \cite{moscardi2021stimuli}. 

In the emerging field of quantum technologies, PCs play a pivotal role in quantum computing and communication, facilitating optical switching and single-photon manipulation \cite{RefA4}. PCs have also proven critical in sensing technologies, enabling the development of ultrasensitive optical sensors, including gas detectors and biosensors, due to their high refractive index sensitivity \cite{RefA3}.

A particularly promising area is the development of tunable photonic materials, where magnetic photonic crystals have emerged as versatile platforms for field-adjustable optical filters and color-shifting materials, with applications in adaptive sensing and camouflage \cite{RefJ4}.

A particularly intriguing development is the use of magnetic nanoparticles in colloidal PCs, where external magnetic fields induce structural periodicity changes, enabling real-time optical tuning \cite{RefJ5,RefJ6}. While prior studies have established that magnetics nanoparticles chain formation governs optical response, a complete understanding of how parameters such as core-shell composition, interparticle distance, and size distribution affect photonic band gaps  remains an active research area \cite{RefJ7, xue2017theoretical}.

In this work, we investigate the optical response of linear arrays of core-shell magnetic nanoparticles using a novel hybrid method combining Mie theory with the Discrete Dipole Approximation (DDA). This approach efficiently incorporates multipolar effects while minimizing computational cost, enabling systematic analysis of structural parameters. Our results align with experimental observations of magnetically tunable reflection in magnetics nanoparticles colloids \cite{RefJ8}, providing a foundation for designing next-generation adaptive photonic devices.

\section{Mie and DDA solving of optical extinction}
\section{\label{sec:1}}
\subsection{Mie theory}
\subsection{\label{sec:2}}
The optical scattering and extinction of non-interacting spherical core-shell NPs can be described by Mie theory as

\begin{equation} Q_{scat}^{Mie} = \frac{2}{k^2b^2} \sum\limits_{n=1}^{\infty} (2n - 1) \left( \left| a_n \right|^2 + \left| b_n \right|^2 \right) \label{ecu:1} \end{equation}

\begin{equation} Q_{ext}^{Mie} = \frac{2}{k^2b^2} \sum\limits_{n=1}^{\infty} (2n - 1) , \text{Re} \left( a_n + b_n \right) \label{ecu:2} \end{equation}

where $k$ is the wavenumber of the incident wave and $a_n$ and $b_n$ are the series coefficients that depend on the internal $a$ and external $b$ radii of the NPs, as well as on the dielectric and magnetic functions of the composing materials \cite{RefB2}. However, in PC-like arrays, particles do present strong interaction with each other and therefore, Mie theory cannot be directly applied. In such cases, an alternative approach is required to accurately describe the optical response of the system.

\subsection{Discrete Dipole Approximation}
\subsection{\label{sec:3}}

Since the PC effect is a collective phenomenon involving a large number of particles with typical sizes above approximately 50 nm, analytically solving its optical response proves to be a highly challenging task. As a result, it is often more practical to employ numerical methods. The optical response of systems composed of non-spherical particles or closely spaced spherical particle arrays can be effectively modeled using the Discrete Dipole Approximation (DDA). In this approach, the system is represented as a spatial distribution of discrete dipoles, each with known polarizability. The electric field at each point is calculated by considering the interaction between the corresponding dipole, the incident electromagnetic wave, and the rest of the dipole distribution. The accuracy of the DDA method, as well as the computational resources required, depends on the size of the discrete dipoles used in the simulation. The maximum allowable dipole size for reliable results depends on the material properties of the nanoparticles and the wavelength range under investigation.

 As demonstrated in \cite{RefJ9}, the scattering and extinction efficiencies in the Discrete Dipole Approximation (DDA) are expressed 
\begin{equation}
Q_{scat}^{dip}=\frac{b^2}{3\pi}k^4\left|\alpha\right|^2
\label{ecu:3}
\end{equation}

\begin{equation}
Q_{ext}^{dip}=\frac{k}{\pi b^2}\text{Im}\left(\alpha\right)
\label{ecu:4}
\end{equation}

where $\alpha$ is the polarizability defined as

\begin{equation}
\nonumber
\alpha=4\pi b^3\frac{\left(\varepsilon_2-\varepsilon_m\right)\left(\varepsilon_1+2\varepsilon_m\right) +h\left(\varepsilon_1-\varepsilon_2\right)\left(\varepsilon_m+2\varepsilon_2\right)}{\left(\varepsilon_2+2\varepsilon_m\right)\left(\varepsilon_1+2\varepsilon_2\right)+h\left(\varepsilon_1-\varepsilon_2\right)\left(2\varepsilon_2-2\varepsilon_m\right)}=\alpha^R+i\alpha^I
\label{ecu:5}
\end{equation}

with $\varepsilon _1$, $\varepsilon _2$  and  $\varepsilon _m$ being the permittivities of core, shell and medium, respectively.
For naked NPs, the expression of the polarizability may be derived from (\ref{ecu:5}) and can be written as

\begin{equation}
\alpha=4\pi b^3\frac{\varepsilon-\varepsilon_m}{\varepsilon+2\varepsilon_m}
\label{ecu:6}
\end{equation}

where $\varepsilon$ is the permittivity of the naked NP. 
 
Studying the optical response of PC systems under continuous-spectrum electromagnetic waves involves significant computational time. Moreover, since the maximum discrete dipole size for DDA is on the order of 10 nm, calculating the complete optical response for particles larger than 50 nm in diameter requires substantial computational resources. Given that even the smallest PC systems are composed of tens of NPs, the direct application of DDA to these systems is computationally prohibitive for standard computers.

The Mie theory on Large Nanoparticles Grating Arrays (MiLaNGA) method addresses these challenges by incorporating an effective polarizability derived from Mie theory into the DDA framework. This approach offers two key advantages: (1) it inherently includes multipolar contributions (unlike classical DDA, which only accounts for dipolar effects), and (2) it significantly reduces computation time.

MiLaNGA provides a powerful tool for analyzing the influence of critical sample parameters such as particle size, structure, and interparticle distance (including variations in size and distance), on the optical response of NP arrays. Furthermore, the method is validated by fitting experimental data, enabling the extraction of structural information about the sample. Its versatility also allows for the prediction of optical behavior in PCs composed of other materials.

\section{Numerical solutions for MNP chains using the MiLaNGa method}
\section{\label{sec:4}}

The novel method proposed in this work for the numerical resolution of linear PC arrays composed of core-shell nanoparticles (CSNPs) involves two fundamental steps. First, the effective polarizability of each individual NP within the array is calculated using Mie theory. Subsequently, the Discrete Dipole Approximation (DDA) is applied to the entire array, incorporating this effective polarizability to numerically determine the theoretical optical response of the PC-like structures. This approach enables the exploration of the optimal conditions for achieving the desired PC tunability, tailored to specific applications.

\subsection{Effective polarizability from Mie theory}
\subsection{\label{sec:5}}

To apply DDA to an assembly of CSNPs, we calculate the extinction and absorption efficiencies of an isolated particle using Mie theory and equate them to the corresponding efficiencies derived from the dipolar approximation for a non-coated particle. This equivalence enables the determination of an effective polarizability ($\alpha_{eff}$), which is subsequently incorporated into the DDA framework.

\begin{eqnarray}
\nonumber
Q_{ext}^{Mie}&=&\frac{2}{k^2b^2}\sum\limits_{n=1}^{\infty}(2n-1)Re\left(a_n+b_n\right)\\
             &=&Q_{ext}^{dip,eff}=\frac{k}{\pi b^2}\alpha_{eff}^{I}
\label{ecu:7}
\end{eqnarray}
 
\begin{eqnarray}
\nonumber
Q_{scat}^{Mie}&=&\frac{2}{k^2b^2}\sum\limits_{n=1}^{\infty}(2n-1)\left( \left|a_n\right|^2+\left|b_n\right|^2 \right)\\
              &=&Q_{scat}^{dip,eff}=\frac{b^2}{3\pi}k^4\left|\alpha_{eff}\right|^2
\label{ecu:8}
\end{eqnarray}

Then, the imaginary and real parts of the effective polarizability are

\begin{equation}
\alpha_{eff}^{I}=\frac{\pi b^2}{k}Q_{ext}^{Mie}
\label{ecu:9}
\end{equation}

\begin{equation}
\alpha_{eff}^{R}=\sqrt{\frac{3\pi }{b^2k^4}Q_{scat}^{Mie}-\left(\alpha_{eff}^{I}\right)^2}
\label{ecu:10}
\end{equation}

\begin{figure}
  \includegraphics{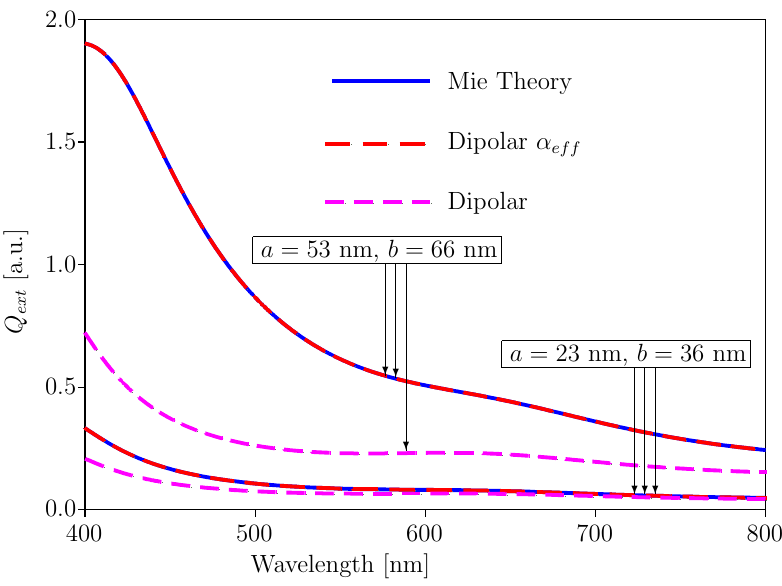}
\caption{\label{fig:1} Comparison between the extinction spectra of a large Fe$_3$O$_4$@SiO$_2$ (53@66 nm and 23@36 nm) CSNP obtained by Mie theory (full blue line), dipolar approximation with effective polarizability (dashed red) and dipolar approximation with polarizability defined by equation 5 (dashed magenta).}
\end{figure}
To illustrate the application of this approach, we employed a model system inspired by the tunable PC developed by Hu \textit{et al.} \cite{RefJ8}. In their work, the authors describe a magnetically assembled one-dimensional chain-like photonic nanostructure composed of (106@13) nm Fe$_3$O$_4$@SiO$_2$ CSNPs.

Figure \ref{fig:1} presents a comparison of the extinction spectra for a large Fe$_3$O$_4$@SiO$_2$ (53@66 nm and 23@36 nm) with dimensions of 53@66 nm and 23@36 nm. The spectra were calculated using three different approaches: Mie theory (solid blue line), the dipolar approximation with effective polarizability (dashed red line), and the dipolar approximation with polarizability defined by equation (\ref{ecu:5})  (dashed magenta line). As anticipated, the blue and red lines are identical by definition, but both deviate from the direct dipolar approximation. This discrepancy arises because the large size of the CSNP  ($b >15$ nm) exceeds the validity limit of the direct dipolar approximation. 

It is important to note that if the discrete dipole approximation (DDA) is applied to an  Fe$_3$O$_4$@SiO$_2$ nanoparticle (NP) with a radius of 23@13 nm—the simplest case in this work—approximately 7400 dipoles are required to obtain a spectrum comparable to that derived from Mie theory. On a desktop computer with 12 GB of RAM and a 3.3 GHz processor, this calculation takes about 6 hours to complete. The proposed method offers a significant advantage in terms of efficiency, as it is substantially faster and considerably less computationally expensive compared to other existing approaches \cite{RefJ18} .  

\begin{figure*}
 \includegraphics{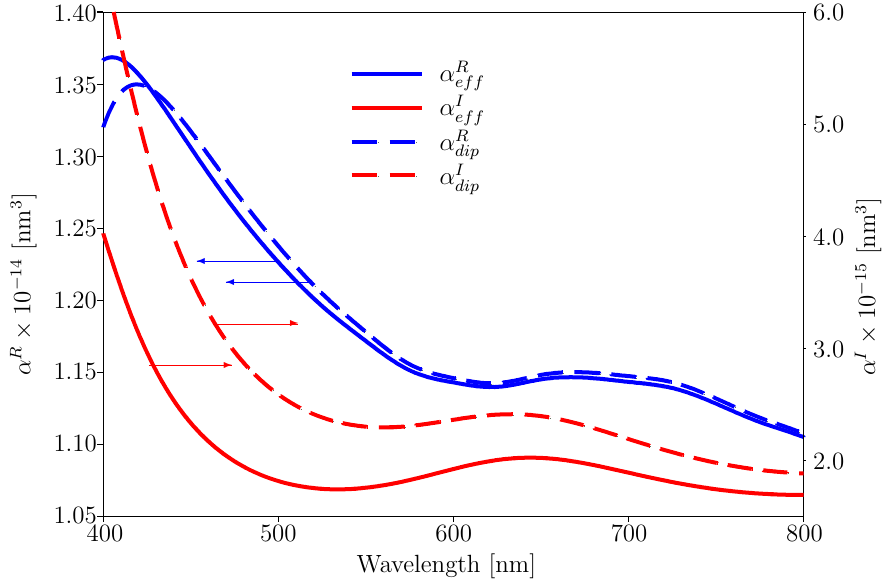}
\caption{\label{fig:2} Real and imaginary effective polarizability from Mie theory compared with their dipolar approximation counterparts.}
\end{figure*}

In Figure \ref{fig:2}, the real and imaginary parts of the effective polarizability derived from Mie theory are compared with those obtained from the dipolar approximation. The noticeable differences observed in the figure arise from the inapplicability of the dipolar approximation within this size-wavelength range.

As the size of the nanoparticles (NPs) increases, the discrepancy between the extinction spectra calculated using the dipolar approximation and Mie theory becomes more pronounced. This is due to the growing significance of multipolar contributions, which cannot be neglected for larger NPs. These differences are particularly evident in the imaginary part of the polarizability, likely because of its direct correlation with absorption processes.

\subsection{DDA solving of MNP chains}
\subsection{\label{sec:6}}

\begin{figure*}
  \includegraphics{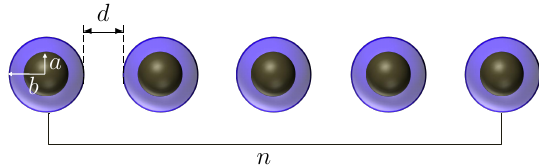}
\caption{\label{fig:3} Scheme of the modeled system. A chain of $n$ spherical core-shell nanoparticles of internal radius $a$ and external radius $b$, separated a distance $d$ from surface-to-surface.}
\end{figure*}
As previously mentioned, the MiLaNGA method enables the study of the dependence of the optical response of photonic crystal (PC)-like arrays on various parameters, as illustrated in Figure  \ref{fig:3}.  To facilitate a direct comparison with the experimental results reported by Hu \textit{et al.} \cite{RefJ8} for magnetic colloidal photonic crystals, we calculated the extinction efficiency for linear chains of Fe$_3$O$_4$@SiO$_2$ core-shell nanoparticles (CSNPs) with an internal@external radius of (53@66) nm, as well as for variations of this system. 

\subsubsection{Radiation mode}
\subsubsection{\label{sec:7}}

\begin{figure*}
\includegraphics{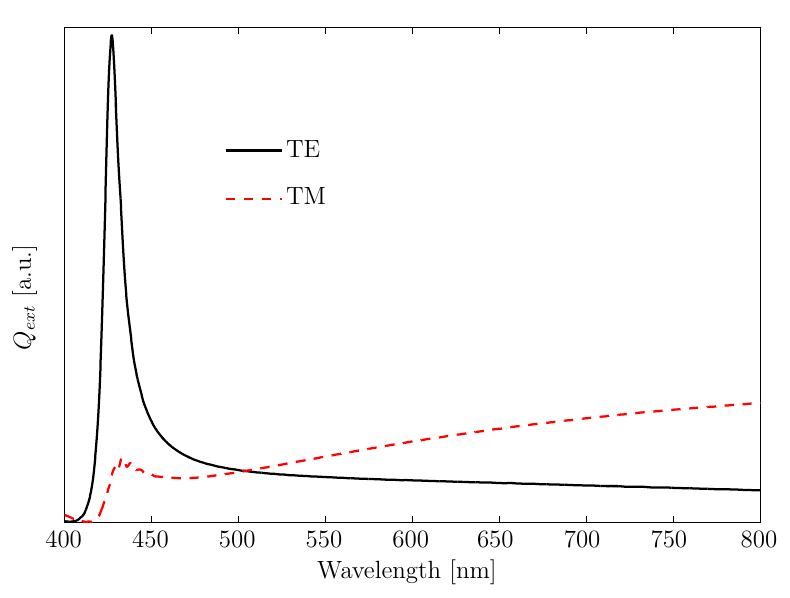}
\caption{\label{fig:4} Comparison between the extinctions of a $a = 53$ nm, $b = 66$ nm, $n = 100$ and $d = 232$ nm chain obtained by considering transverse electric (TE) and transverse magnetic (TM) incidence modes.}
\end{figure*}
Since the photonic crystal (PC) behavior arises from geometrical effects, the electromagnetic mode of incidence plays a crucial role in determining the optical response of 2D arrays. Figure \ref{fig:4} compares the extinction spectra of a chain of $n= 100$ nanoparticles (NPs), where $n$ represents the number of NPs in the chain, with a surface-to-surface separation of $d= 100$ nm. The spectra were calculated for both transverse electric (TE) and transverse magnetic (TM) modes. The results clearly indicate that the selective extinction responsible for the PC effect is predominantly driven by TE mode incidence. Consequently, all subsequent simulations presented in this work will focus exclusively on TE mode incidence. 

\subsubsection{Particle number}
\subsubsection{\label{sec:8}}

The number of core-shell nanoparticles (CSNPs) in the array, denoted as $n$, is typically a large statistical value in real samples. However, for simulations, increasing the number of particles significantly extends the computation time. Therefore, it is essential to demonstrate that the calculation results are independent of the particle number, especially since the simulations presented here will focus on linear chains of up to 100 particles.

\begin{figure*}
\includegraphics{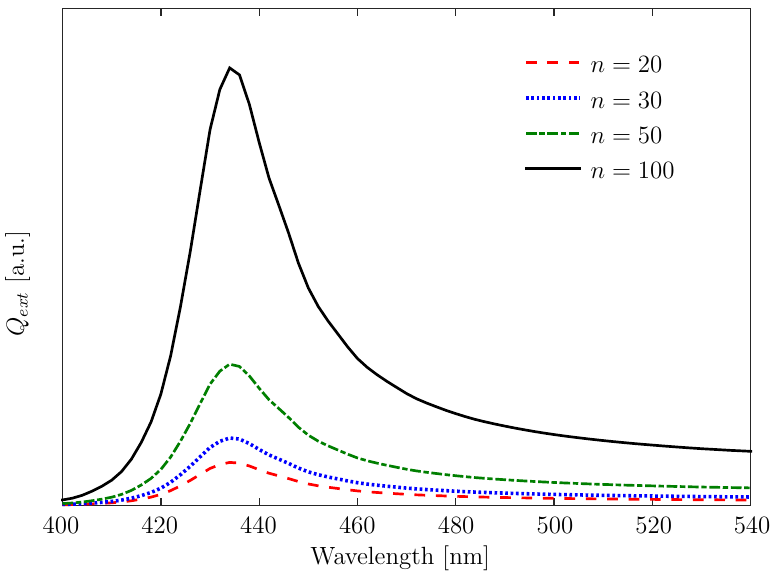}
\caption{\label{fig:5} Optical extinction efficiency for chains of Fe$_3$O$_4$@SiO$_2$ NPs with $a = 53$ nm, $b = 66$ nm and $d = 232$ nm, for several n values. While the height of the peak increase with $n$, the maximum position shift is negligible.}
\end{figure*}

Figure \ref{fig:5} illustrates the dependence of the extinction efficiency peak on the number of CSNPs in the linear chain. As the number of particles increases, the height of the extinction peak grows, while the peak's wavelength shift remains negligible.

These results allow us to conclude that the predictions from the simulations are applicable to real systems, where the number of particles is several orders of magnitude larger.

\subsubsection{Core size and shell thickness}
\subsubsection{\label{sec:9}}

The dimensions of the core and shell are fundamental parameters of the core-shell nanoparticles (CSNPs) that can be precisely controlled during synthesis to achieve specific values tailored for the desired optical response. In this section, we analyze the impact of these parameters on the optical response by examining their effects in three distinct ways. 

\begin{figure*}
\includegraphics{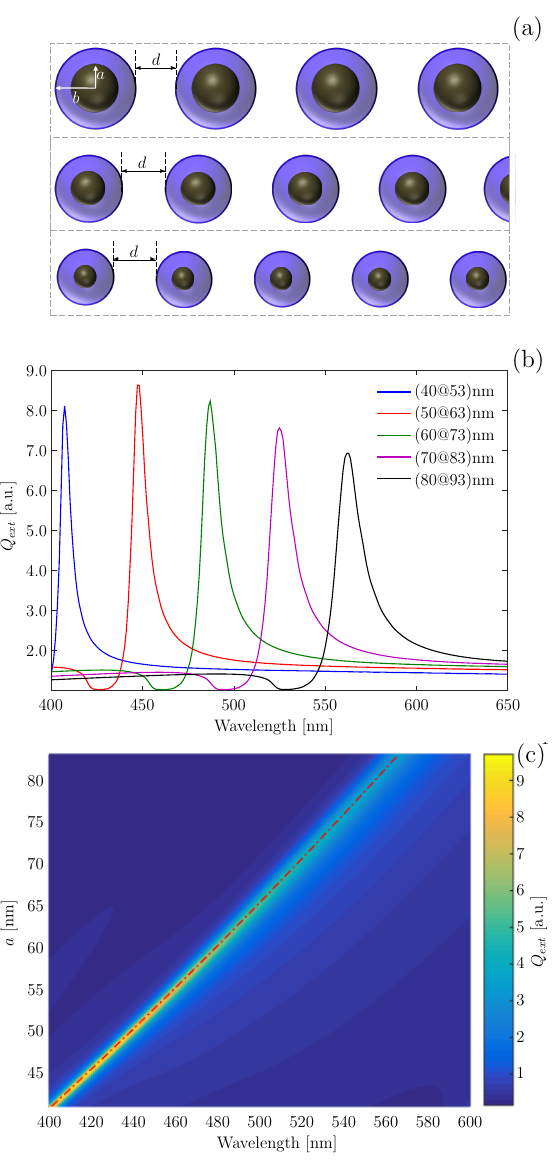}
\caption{\label{fig:6} (a) Progression of the simulations. Optical extinction efficiency $Q_{ext}$ is calculated for different values of $a$ while maintaining the values of $b$ and $d$ constant. (b) $Q_{ext}$ versus wavelength for $b-a =$ 13 nm, $d =$ 120 nm and several values of $a$. (c) $Q_{ext}$ versus wavelength and core size $a$ for $b-a =$ 13 nm, $d =$ 120 nm.}
\end{figure*}

Figure \ref{fig:6}-b displays the extinction efficiency for various values of the core radius $a$ in Fe$_3$O$_4@$SiO$_2$ core-shell nanoparticles (CSNPs) arranged in a chain of 100 particles with a surface-to-surface separation of $d=120$ nm. The shell thickness, defined as $b-a$, is kept constant at 13 nm. This scenario represents a common synthesis process for CSNPs, where cores of different sizes are coated with a 13 nm thick shell, simulating sequential shell growth on varying core dimensions. 

Figure \ref{fig:6}-c presents a color map of the extinction efficiency $Q_{ext}$ as a function of the core radius $a$ and the wavelength $\lambda$. The map reveals that the peak maximum shifts to longer wavelengths as $a$ increases. A minimum core size of $a=41$ nm is required for the peak to fall within the visible region. The dependence of the peak position on the core radius was fitted with a second-degree polynomial, yielding an almost linear relationship described by $\lambda_p= -8.1\times 10^{-3} a^2+4.92 a+ 2.13 \times 10^2$. 

Having examined the effects of varying the core radius of the nanoparticles (NPs) while maintaining a constant shell thickness, it is equally crucial to explore the impact of altering the shell thickness while keeping the core radius constant. 

\begin{figure*}
\includegraphics[width=0.70\textwidth]{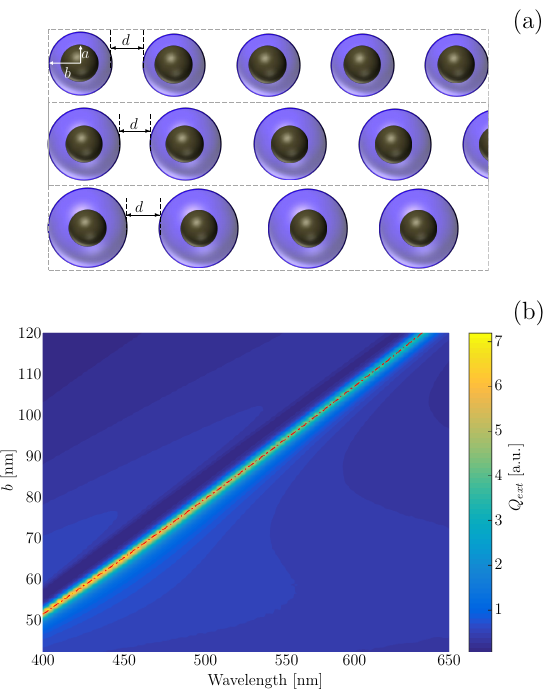}
\caption{\label{fig:7} (a) Progression of the simulations. Optical extinction efficiency $Q_{ext}$ is calculated for different values of $b$ while maintaining the values of $b-a$ and $d$. (b) $Q_{ext}$ versus wavelength and core size $a$ for $b-a =$ 13 nm, $d =$ 120 nm.}
\end{figure*}

Figure \ref{fig:7} illustrates the variations in the extinction efficiency $Q_{ext}$ as the shell thickness $b-a$ is changed, while keeping the core radius constant at $a= 40 $. The dependence of the peak position on the external radius $b$ was fitted with a second-degree polynomial, revealing an almost linear relationship described by $\lambda_p=9.477\times 10^{-5} b^2+3.398 b-224.7$.
 
In the previous cases, the size of the nanoparticle (NP) was increased by either enlarging the shell thickness or the core radius. However, an interesting scenario to analyze arises when the ratio between the core radius and the shell thickness is altered without changing the overall size of the NP. This situation provides valuable insights into how the internal material distribution (core and shell) influences the optical properties, while keeping the total dimensions of the NP constant. 

\begin{figure*}
\includegraphics[width=0.70\textwidth]{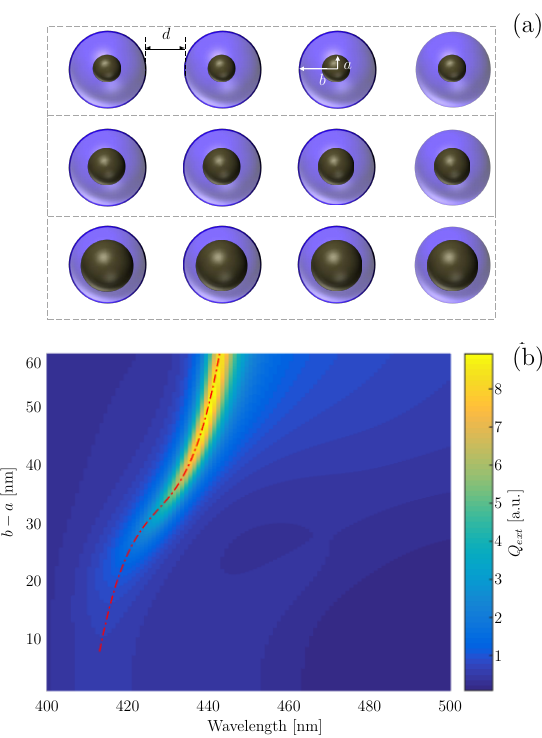}
\caption{\label{fig:8} (a) Progression of the simulations. Optical extinction efficiency $Q_{ext}$ is calculated for different values of $b-a$ while maintaining the values of $b$ and $d$ constant. (b) $Q_{ext}$ versus wavelength and shell thickness a for $b =$ 65 nm, $d =$ 120 nm.}
\end{figure*}

Figure  \ref{fig:8} illustrates the variation of the extinction efficiency ($Q_{ext}$ ) as a function of the core radius ($a$), while keeping the total particle size ($b$) constant. The peak position ($\lambda_p$) exhibits a weak cubic dependence on $a$, described by:

$$\lambda_p=-4.00\times 10^{-4} a^3+4.30\times 10^{-2} a^2- 5.70\times 10^{-1} a+4.14 \times 10^2$$

\subsubsection{Particle composition}
\subsubsection{\label{sec:10}}

Thus far, we have examined how the peak position varies with the dimensions of the nanostructures. However, the optical response of the system is also critically influenced by the composition of the core-shell nanoparticles (CSNPs), as their dielectric functions dictate light-matter interactions. By selecting an appropriate shell material, the extinction peak position can be further tuned, offering additional control over the system's spectral properties.

\begin{figure*}
\includegraphics{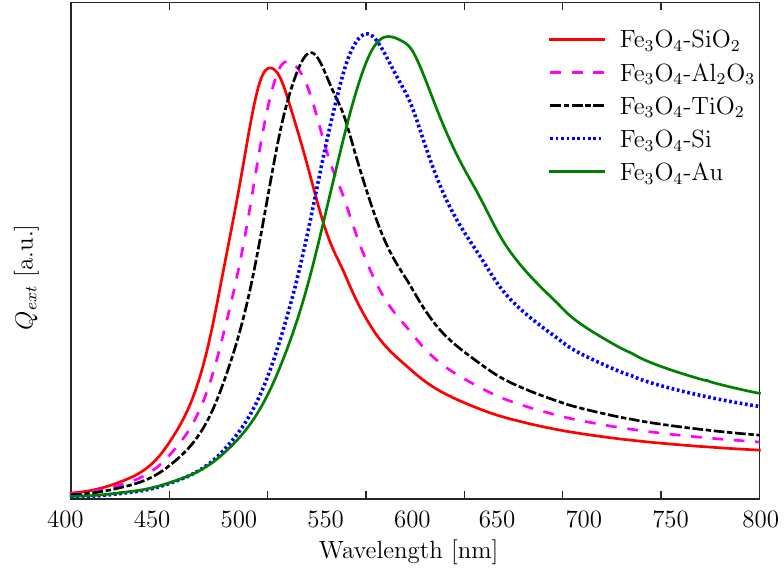}
\caption{\label{fig:9} Optical extinction efficiency $Q_{ext}$ for different shell materials for $a=53$ nm, $b =$ 63 nm and $d =120$ nm.}
\end{figure*}

Figure \ref{fig:9} presents the extinction spectra for a chain of magnetite-core nanoparticles with a fixed core radius $a = 53 $ nm and a shell radius $b = 63 $ nm, coated with different materials: silicon dioxide (SiO$_2$), alumina (Al$_2$O$_3$), titanium dioxide (TiO$_2$), silicon (Si), and gold (Au). The surface-to-surface distance between particles is maintained at $d = 120 $ nm.

The results demonstrate that the choice of shell material significantly influences the peak position in the extinction spectra. This effect arises from variations in the shell's dielectric properties, which alter the effective polarizability of the nanoparticle. Notably, if needed, a redshift of the extinction peak can be counterbalanced by reducing the particle size.

\subsubsection{Interparticle separation}
\subsubsection{\label{sec:11}}

The interparticle separation in our photonic crystal is a critical parameter, as it governs the system's response to external magnetic fields. The magnetite cores enable direct manipulation of the crystal structure through applied magnetic fields - by adjusting the distance between the external magnet and the sample, we can control both the magnetic interaction strength and the interparticle spacing. This tunability directly influences the photonic properties of the system, manifesting as measurable shifts in the reflection peak position \textbackslash{}cite\{RefJ5\}.  

\begin{figure*}
\includegraphics{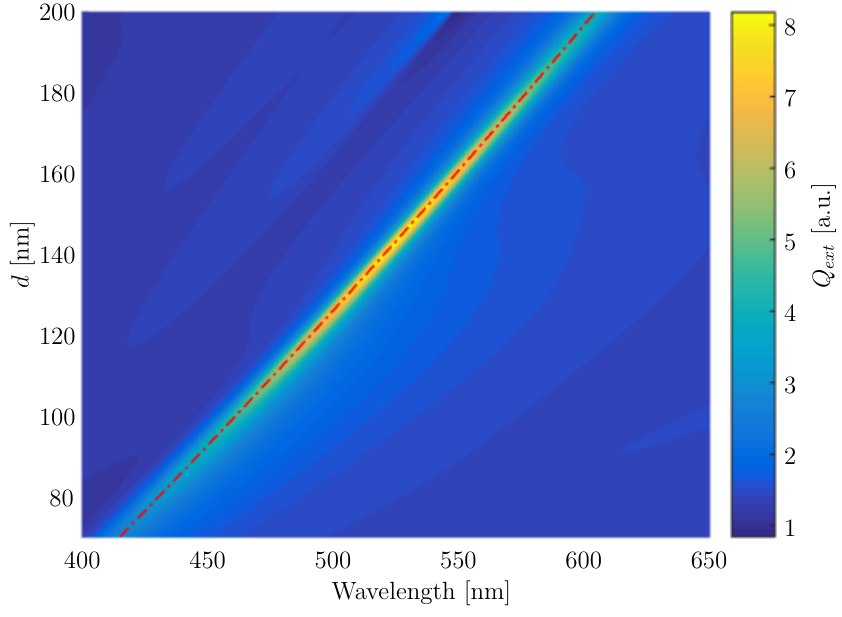}
\caption{\label{fig:10} Optical extinction efficiency $Q_{ext}$ versus interparticle separation for $a = 53$ nm and $b =$ 66 nm.}
\end{figure*}

The model presented in this work can easily reproduce this behavior. Figure \ref{fig:10} shows the evolution in the extinction peak position of a 53@66 nm CSNP chain of Fe$_3$O$_4$@SiO$_2$ as a function of the interparticle distance $d\in$[70,200] nm. As reported by Hu \textit{et al.} \cite{RefJ8} for the reflection peak, the maximum shifts to longer wavelengths with increasing interparticle distance, maintaining a practically linear dependency $\lambda_p= -8.00\times 10^{-4} d^2 +1.67 d+302$ in the visible region .

In the case of arrays of higher dimensionality, the optical behavior responds to an ``effective interparticle distance'' depending on the angle between the incidence and observation directions.

\subsubsection{Particle size distribution}
\subsubsection{\label{sec:12}}

\begin{figure*}
\includegraphics{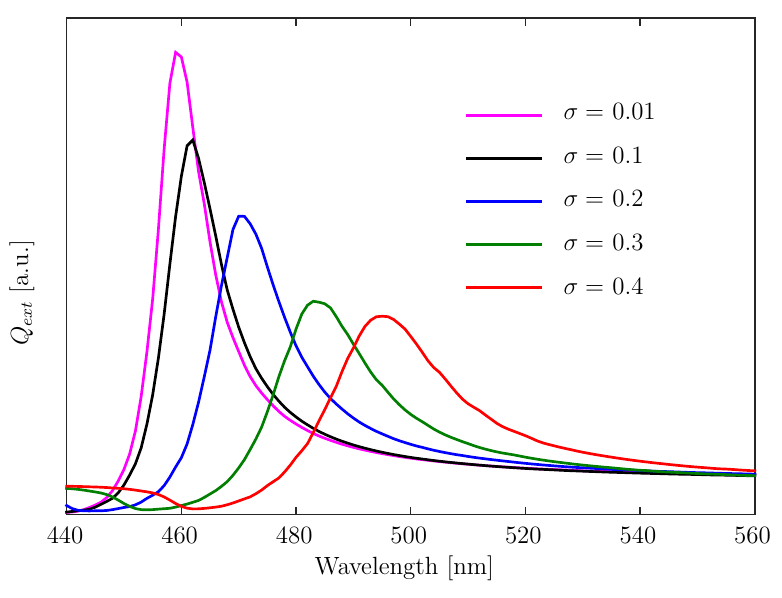}
\caption{\label{fig:11} (a) Example of log-normal distributions used in the simulations. (b) Optical extinction efficiency $Q_{ext}$ for different particle size distribution for $a = 50$ nm, $b =$ 63 nm and $d = 120$ nm.}
\end{figure*}

All results presented so far correspond to a monosized system. In real samples, size dispersion is always present and affects the overall optical response.
Figure \ref{fig:11}-b shows the extinction peaks of 50@63 nm Fe$_3$O$_4$@SiO$_2$ CSNP chains with different log-normal size standard deviations $\sigma$ depicted in figure \ref{fig:11}-a. The main effect of a larger size dispersion is a redshift of the maximum and a broadening of the peak.

\subsubsection{Interparticle distance dispersion }
\subsubsection{\label{sec:13}}

The force on the CSNP due to the external magnetic field is proportional to the magnetic moment of the particles and to the gradient of the field. At the same time, the magnetic moment of the particles grows with field strength. On the other hand, the spatial variation of the field gradient is bigger for smaller distances to the magnet but, for larger distances, the other forces involved in particle separation are comparable to the effect of the magnet. Being the additional forces more randomly distributed, the dependence of interparticle distance dispersion with magnet-to-sample distance has not a general behaviour and must be determined in each particular case. 

\begin{figure*}
\includegraphics{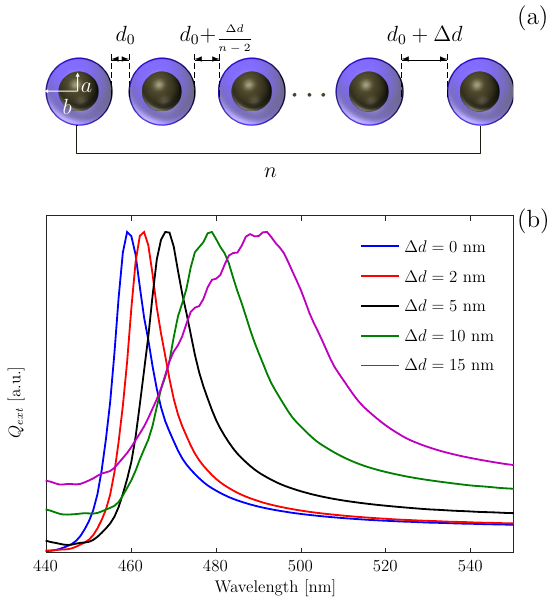}
\caption{\label{fig:12} Optical extinction efficiency $Q_{ext}$ for different interparticle distance dispersion for $a=50$ nm, $b=$ 63 nm and $d_0=120$ nm.}
\end{figure*}

Figure \ref{fig:12} shows the effect  of interparticle distance dispersion $\Delta d$ on the extinction. The interparticle distance change linearly through the array, from an initial value $d_0$ to a final separation of $d_0+\Delta d$ between the last pair of particles. For a larger $\Delta d$, the extinction peak is broader and shifts to larger wavelengths. 

By leveraging our understanding of how photonic crystal parameters affect both the spectral position and linewidth of the resonance peaks, we can perform quantitative comparisons between our theoretical predictions and experimental measurements. 

XXXXXXXXXXXXXXXXXXXXXXXXXXXXXXXXXXXXXXXX
\section{Experimental data fitting}
\section{\label{sec:14}}

As a validation of the calculated results, in this section we will fit the experimental spectra presented by Hu \textit{et al.} \cite{RefJ8}, taking into account all the features considered in the previous section.

\begin{figure*}
\includegraphics{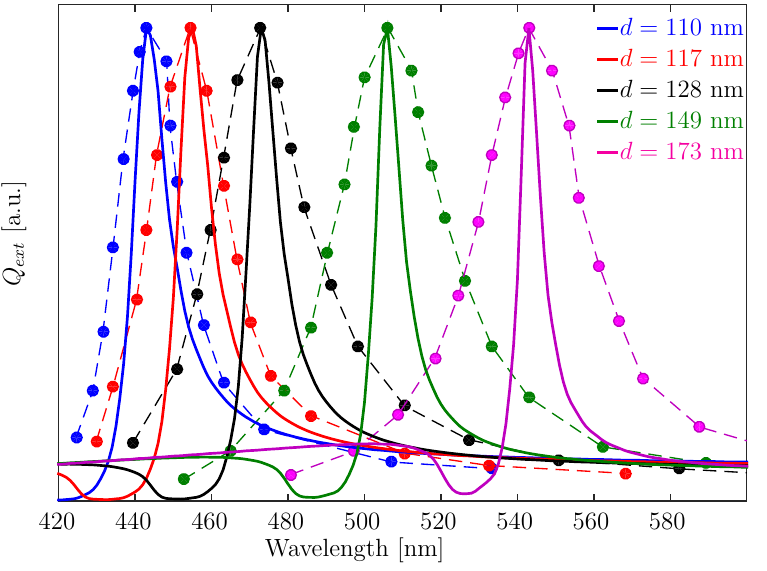}
\caption{\label{fig:13} Comparison between experimental spectra and the corresponding spectra calculated for a monosized sample with zero dispersion in interparticle distance.}
\end{figure*}

To a first approximation, we establish that each experimental spectral peak position corresponds to a unique interparticle spacing, following the relationship derived in Section \ref{sec:11}. Figure \ref{fig:13} presents a direct comparison between the experimental spectra reported by Hu et al. \cite{RefJ8} and our theoretical simulations, where the system is modeled with monodisperse nanoparticles and a uniform interparticle separation distance \textit{d}. The analysis reveals a systematic discrepancy, with experimental resonance peaks showing significantly broader linewidths compared to theoretical predictions, suggesting the presence of unaccounted for heterogeneity in the experimental system.

\begin{figure*}
\includegraphics{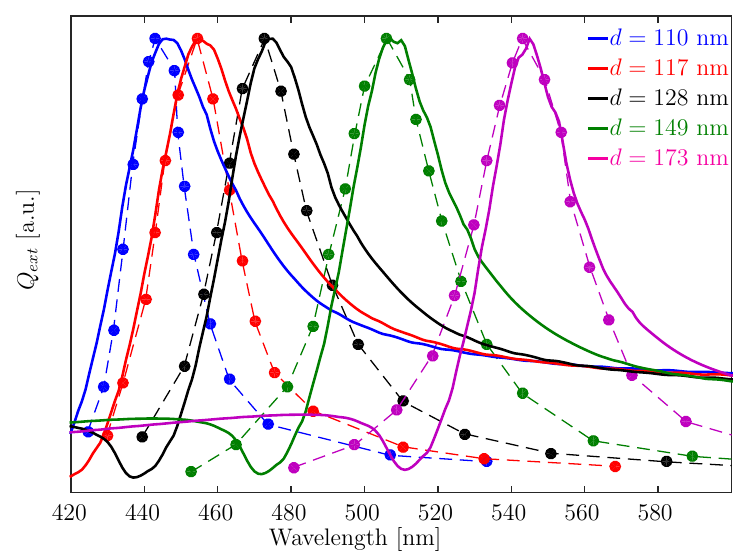}
\caption{\label{fig:14} Comparison between experimental spectra and the corresponding spectra calculated for a log-normal size distribution sample with zero dispersion in interparticle distance. }
\end{figure*}

The fitting of the experimental data can be improved then by considering the effects of particle size dispersion mentioned in section \ref{sec:9}. Figure \ref{fig:14} shows the comparison between the same experimental spectra and the corresponding calculation, now assuming a log-normal core size distribution with $\sigma=0.5$ common to all spectra and correcting $d$ value for each case. The correspondence improves for larger interparticle distance $d$ (hence, larger peak wavelength), but the calculated spectra for smaller $d$ still differs considerably from the experiment.

\begin{figure*}
\includegraphics{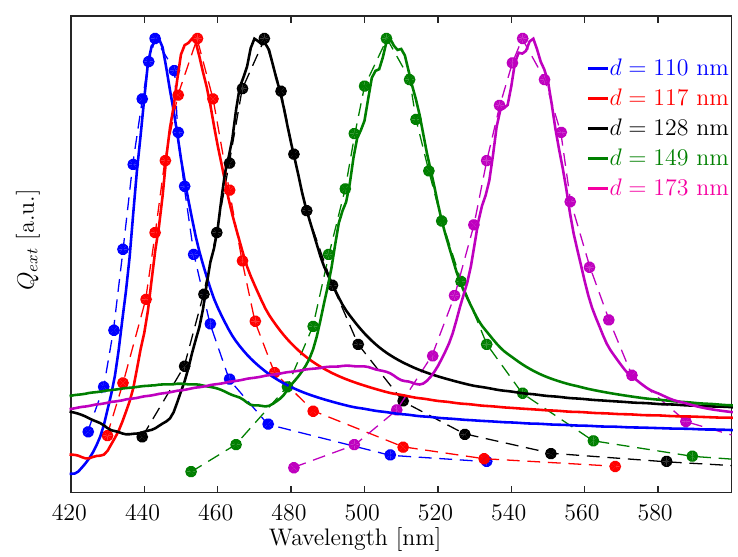}
\caption{\label{fig:15} Comparison between experimental spectra and the corresponding spectra calculated for a log-normal size distribution sample considering a different interparticle distance dispersion for each spectrum..}
\end{figure*}

Finally, experimental data is fitted again adding an interparticle distance dispersion parameter $\Delta d$. Figure \ref{fig:15} shows a very good agreement between calculation and experiment, indicating a growing dispersion with larger interparticle distance. The final results indicate a mean core size of $<a>=56 $ nm, and a standard deviation $SD=20 $ nm.    

\begin{figure*}
\includegraphics{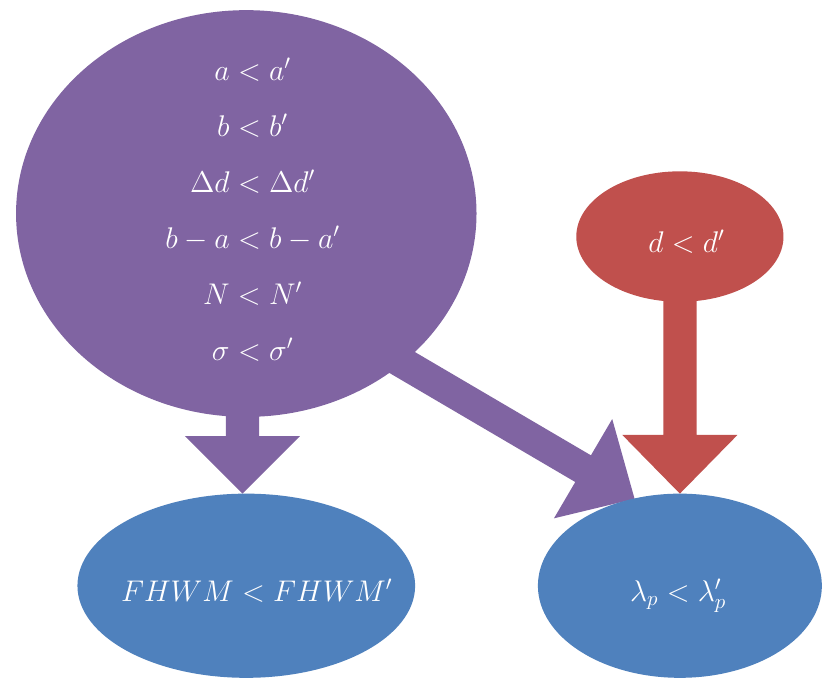}
\caption{\label{fig:16} Summary of the effects of the NPs array parameters on the extinction peak of the photonic crystal.}
\end{figure*}

\section{Discussion and conclusions}
\section{\label{sec:15}}

The photonic crystals behavior of linear arrays of magnetic core-dielectric shell nanoparticles (CSNP) was studied by a novel method combining Mie theory with DDA. This hybrid technique allows including  multipolar contributions effects as well as reducing calculation time. The method was validated by reproducing the general qualitative behaviour of all experimental results reported in the literature and additionally, by fitting with very good agreement the particular results analyzed.   

Extinction spectra for an initial linear array of $n = 100$ Fe$_3$O$_4$@SiO$_2$ CSNPs with (53@66) nm internal@ external radius and a surface-to-surface separation of 100 nm, mimicking the systems studied by Hu \textit{et al.} \cite{RefJ8} were calculated in  Section \ref{sec:4}, where also the effect on the extinction efficiency of each of the system's parameters was studied separately.

By comparison of the corresponding spectra, it was clearly demonstrated that the main contribution for the PC effect of anomalous extinction comes from the transverse electric radiation mode. Hence, all the subsequent spectra were calculated considering only this propagation mode. 

Since the maximum number of particles considered in the calculations is far below the statistical number of particles in a real sample, the dependence of the extinction spectra on the number of particles was studied. The results indicate that the position of the extinction maximum and the peak width do not depend on the number of particles in the array, being the spectra affected only in the intensity of the peaks. This allows us to consider the results of the calculation as representative of the real systems.

The influence of the CSNPs structure was studied.  In all cases, enlarging particle dimensions, such as core radius, shell thickness and external radius, leads to a redshift of the reflection peak and an increase of its $FWHM$. A similar effect is also observed for increasing values of shell refractive index. 
In both cases, this behaviour is due to an overall change in the effective refractive index distribution, which, in turn, modifies its contrast and allows geometry-based reflection tunability. Besides, as expected, increasing interparticle distance leads to a redshift of reflection peak due to an enlargement of the refractive index periodicity.

 Finally, the effect of particle size distribution $\sigma$ and interparticle distance dispersion $\Delta d$ was considered.  An increase in the former shifts the peak to larger wavelengths and enlarges the $FWHM$, while considering an interparticle distance dispersion, leads to a similar qualitative effect. All these results are summarized in figure \ref{fig:16}.

It is important to notice that the use of surface-to-surface distance (instead of center-to-center) as a fitting parameter, causes the modification of the effective refractive index together with the center-to-center distance. So, the effects of these two parameters are combined in the simulations.  

In order to reproduce the experimental results presented by Hu \textit{et al.}, it was necessary to consider a log-normal distribution of sizes and a dispersion of the interparticle distance. Including these features, it was possible to achieve a very good agreement between the calculated curves and the experimental data, thus obtaining previously unabailable information of the system. A log-normal distribution of mean radius $<a>=60 $ nm and the standard deviation $20 $ nm was obtained, which couldn't be compared with the source since the authors didn't report it. On the other hand, our method has the advantage of retrieving interparticle mean distance as well as its dispersion, which are parameters not easily obtainable via direct measurements. 

The MiLaNGA method presented in this work has been succesfully tested solving and simulating a PC like system. Its use allows to predict the optical behaviour of periodic arrays of single and multilayer NPs of any composition, even considering size distribution and interparticle distance dispersion. Additionaly, MiLaNGA can be used as an experimental data fitting tool to  extract unknown system information from extinction spectra. All these features constitute an accessible and easy to use resource for the understanding, designing and characterization of tunable photonic crystals of any kind, which constitute the future of this promising branch of photonic technology.

\section{acknowledgements}
This work was granted by PIP 0280 of CONICET, PME2006-00018
of ANPCyT, grant 11/I197 of Facultad de Ingeniería, UNLP. D. C.
Schinca is Member of Comisión de Investigaciones Científicas de
la Provincia de Buenos Aires (CICBA), Argentina. L. B. Scaffardi, is
researcher of CONICET.

\bibliography{Biblio}

\begin{thebibliography}{22}%
\makeatletter
\providecommand \@ifxundefined [1]{%
 \@ifx{#1\undefined}
}%
\providecommand \@ifnum [1]{%
 \ifnum #1\expandafter \@firstoftwo
 \else \expandafter \@secondoftwo
 \fi
}%
\providecommand \@ifx [1]{%
 \ifx #1\expandafter \@firstoftwo
 \else \expandafter \@secondoftwo
 \fi
}%
\providecommand \natexlab [1]{#1}%
\providecommand \enquote  [1]{``#1''}%
\providecommand \bibnamefont  [1]{#1}%
\providecommand \bibfnamefont [1]{#1}%
\providecommand \citenamefont [1]{#1}%
\providecommand \href@noop [0]{\@secondoftwo}%
\providecommand \href [0]{\begingroup \@sanitize@url \@href}%
\providecommand \@href[1]{\@@startlink{#1}\@@href}%
\providecommand \@@href[1]{\endgroup#1\@@endlink}%
\providecommand \@sanitize@url [0]{\catcode `\\12\catcode `\$12\catcode
  `\&12\catcode `\#12\catcode `\^12\catcode `\_12\catcode `\%12\relax}%
\providecommand \@@startlink[1]{}%
\providecommand \@@endlink[0]{}%
\providecommand \url  [0]{\begingroup\@sanitize@url \@url }%
\providecommand \@url [1]{\endgroup\@href {#1}{\urlprefix }}%
\providecommand \urlprefix  [0]{URL }%
\providecommand \Eprint [0]{\href }%
\providecommand \doibase [0]{http://dx.doi.org/}%
\providecommand \selectlanguage [0]{\@gobble}%
\providecommand \bibinfo  [0]{\@secondoftwo}%
\providecommand \bibfield  [0]{\@secondoftwo}%
\providecommand \translation [1]{[#1]}%
\providecommand \BibitemOpen [0]{}%
\providecommand \bibitemStop [0]{}%
\providecommand \bibitemNoStop [0]{.\EOS\space}%
\providecommand \EOS [0]{\spacefactor3000\relax}%
\providecommand \BibitemShut  [1]{\csname bibitem#1\endcsname}%
\let\auto@bib@innerbib\@empty
\bibitem [{\citenamefont {Joannopoulos}\ \emph {et~al.}(2011)\citenamefont
  {Joannopoulos}, \citenamefont {Johnson}, \citenamefont {Winn},\ and\
  \citenamefont {Meade}}]{RefB1}%
  \BibitemOpen
  \bibfield  {author} {\bibinfo {author} {\bibfnamefont {J.~D.}\ \bibnamefont
  {Joannopoulos}}, \bibinfo {author} {\bibfnamefont {S.~G.}\ \bibnamefont
  {Johnson}}, \bibinfo {author} {\bibfnamefont {J.~N.}\ \bibnamefont {Winn}}, \
  and\ \bibinfo {author} {\bibfnamefont {R.~D.}\ \bibnamefont {Meade}},\
  }\href@noop {} {\emph {\bibinfo {title} {Photonic crystals: molding the flow
  of light}}},\ \bibinfo {edition} {1st}\ ed.\ (\bibinfo  {publisher}
  {Princeton university press.},\ \bibinfo {year} {2011})\BibitemShut {NoStop}%
\bibitem [{\citenamefont {Yablonovitch}(1987)}]{RefY1}%
  \BibitemOpen
  \bibfield  {author} {\bibinfo {author} {\bibfnamefont {E.}~\bibnamefont
  {Yablonovitch}},\ }\href@noop {} {\bibfield  {journal} {\bibinfo  {journal}
  {Physical Review Letters}\ }\textbf {\bibinfo {volume} {58}},\ \bibinfo
  {pages} {2059} (\bibinfo {year} {1987})}\BibitemShut {NoStop}%
\bibitem [{\citenamefont {Wang}\ \emph {et~al.}(2023)\citenamefont {Wang},
  \citenamefont {Dong}, \citenamefont {Bu},\ and\ \citenamefont
  {Wang}}]{wang2023design}%
  \BibitemOpen
  \bibfield  {author} {\bibinfo {author} {\bibfnamefont {K.}~\bibnamefont
  {Wang}}, \bibinfo {author} {\bibfnamefont {X.}~\bibnamefont {Dong}}, \bibinfo
  {author} {\bibfnamefont {Y.}~\bibnamefont {Bu}}, \ and\ \bibinfo {author}
  {\bibfnamefont {X.}~\bibnamefont {Wang}},\ }\href@noop {} {\bibfield
  {journal} {\bibinfo  {journal} {Journal of the American Ceramic Society}\
  }\textbf {\bibinfo {volume} {106}},\ \bibinfo {pages} {7146} (\bibinfo {year}
  {2023})}\BibitemShut {NoStop}%
\bibitem [{\citenamefont {Lonergan}\ and\ \citenamefont
  {O'Dwyer}(2023)}]{lonergan2023many}%
  \BibitemOpen
  \bibfield  {author} {\bibinfo {author} {\bibfnamefont {A.}~\bibnamefont
  {Lonergan}}\ and\ \bibinfo {author} {\bibfnamefont {C.}~\bibnamefont
  {O'Dwyer}},\ }\href@noop {} {\bibfield  {journal} {\bibinfo  {journal}
  {Advanced Materials Technologies}\ }\textbf {\bibinfo {volume} {8}},\
  \bibinfo {pages} {2201410} (\bibinfo {year} {2023})}\BibitemShut {NoStop}%
\bibitem [{\citenamefont {Chaudhary}\ \emph {et~al.}(2022)\citenamefont
  {Chaudhary}, \citenamefont {Kumar}, \citenamefont {Pandey},\ and\
  \citenamefont {Kumar}}]{chaudhary2022advances}%
  \BibitemOpen
  \bibfield  {author} {\bibinfo {author} {\bibfnamefont {V.~S.}\ \bibnamefont
  {Chaudhary}}, \bibinfo {author} {\bibfnamefont {D.}~\bibnamefont {Kumar}},
  \bibinfo {author} {\bibfnamefont {B.~P.}\ \bibnamefont {Pandey}}, \ and\
  \bibinfo {author} {\bibfnamefont {S.}~\bibnamefont {Kumar}},\ }\href@noop {}
  {\bibfield  {journal} {\bibinfo  {journal} {IEEE sensors journal}\ }\textbf
  {\bibinfo {volume} {23}},\ \bibinfo {pages} {1012} (\bibinfo {year}
  {2022})}\BibitemShut {NoStop}%
\bibitem [{\citenamefont {Tran}\ \emph {et~al.}(2022)\citenamefont {Tran},
  \citenamefont {Kim}, \citenamefont {Oh}, \citenamefont {Jeong},\ and\
  \citenamefont {Lee}}]{RefT2}%
  \BibitemOpen
  \bibfield  {author} {\bibinfo {author} {\bibfnamefont {V.~T.}\ \bibnamefont
  {Tran}}, \bibinfo {author} {\bibfnamefont {J.}~\bibnamefont {Kim}}, \bibinfo
  {author} {\bibfnamefont {S.}~\bibnamefont {Oh}}, \bibinfo {author}
  {\bibfnamefont {K.-J.}\ \bibnamefont {Jeong}}, \ and\ \bibinfo {author}
  {\bibfnamefont {J.}~\bibnamefont {Lee}},\ }\href@noop {} {\bibfield
  {journal} {\bibinfo  {journal} {Small}\ }\textbf {\bibinfo {volume} {18}},\
  \bibinfo {pages} {2200317} (\bibinfo {year} {2022})}\BibitemShut {NoStop}%
\bibitem [{\citenamefont {Yang}\ \emph {et~al.}(2023)\citenamefont {Yang},
  \citenamefont {band Ranran~Ma}, \citenamefont {Wu}, \citenamefont {Chen},
  \citenamefont {Zhang}, \citenamefont {Ye}, \citenamefont {You},\ and\
  \citenamefont {Xiao}}]{RefT3}%
  \BibitemOpen
  \bibfield  {author} {\bibinfo {author} {\bibfnamefont {S.}~\bibnamefont
  {Yang}}, \bibinfo {author} {\bibfnamefont {R.~D.}\ \bibnamefont {band
  Ranran~Ma}}, \bibinfo {author} {\bibfnamefont {M.}~\bibnamefont {Wu}},
  \bibinfo {author} {\bibfnamefont {P.}~\bibnamefont {Chen}}, \bibinfo {author}
  {\bibfnamefont {Y.}~\bibnamefont {Zhang}}, \bibinfo {author} {\bibfnamefont
  {A.}~\bibnamefont {Ye}}, \bibinfo {author} {\bibfnamefont {L.}~\bibnamefont
  {You}}, \ and\ \bibinfo {author} {\bibfnamefont {D.}~\bibnamefont {Xiao}},\
  }\href@noop {} {\bibfield  {journal} {\bibinfo  {journal} {Journal of
  Magnetism and Magnetic Materials}\ }\textbf {\bibinfo {volume} {585}},\
  \bibinfo {pages} {171097} (\bibinfo {year} {2023})}\BibitemShut {NoStop}%
\bibitem [{\citenamefont {Noda}\ \emph {et~al.}(2000)\citenamefont {Noda},
  \citenamefont {Tomoda}, \citenamefont {Yamamoto},\ and\ \citenamefont
  {Chutinan}}]{RefT4}%
  \BibitemOpen
  \bibfield  {author} {\bibinfo {author} {\bibfnamefont {S.}~\bibnamefont
  {Noda}}, \bibinfo {author} {\bibfnamefont {K.}~\bibnamefont {Tomoda}},
  \bibinfo {author} {\bibfnamefont {N.}~\bibnamefont {Yamamoto}}, \ and\
  \bibinfo {author} {\bibfnamefont {A.}~\bibnamefont {Chutinan}},\ }\href@noop
  {} {\bibfield  {journal} {\bibinfo  {journal} {Science}\ }\textbf {\bibinfo
  {volume} {289}},\ \bibinfo {pages} {604} (\bibinfo {year}
  {2000})}\BibitemShut {NoStop}%
\bibitem [{\citenamefont {Cai}\ \emph {et~al.}(2021)\citenamefont {Cai},
  \citenamefont {Li}, \citenamefont {Ravaine}, \citenamefont {He},
  \citenamefont {Song}, \citenamefont {Yin}, \citenamefont {Zheng},
  \citenamefont {Teng},\ and\ \citenamefont {Zhang}}]{cai2021colloidal}%
  \BibitemOpen
  \bibfield  {author} {\bibinfo {author} {\bibfnamefont {Z.}~\bibnamefont
  {Cai}}, \bibinfo {author} {\bibfnamefont {Z.}~\bibnamefont {Li}}, \bibinfo
  {author} {\bibfnamefont {S.}~\bibnamefont {Ravaine}}, \bibinfo {author}
  {\bibfnamefont {M.}~\bibnamefont {He}}, \bibinfo {author} {\bibfnamefont
  {Y.}~\bibnamefont {Song}}, \bibinfo {author} {\bibfnamefont {Y.}~\bibnamefont
  {Yin}}, \bibinfo {author} {\bibfnamefont {H.}~\bibnamefont {Zheng}}, \bibinfo
  {author} {\bibfnamefont {J.}~\bibnamefont {Teng}}, \ and\ \bibinfo {author}
  {\bibfnamefont {A.}~\bibnamefont {Zhang}},\ }\href@noop {} {\bibfield
  {journal} {\bibinfo  {journal} {Chemical Society Reviews}\ }\textbf {\bibinfo
  {volume} {50}},\ \bibinfo {pages} {5898} (\bibinfo {year}
  {2021})}\BibitemShut {NoStop}%
\bibitem [{\citenamefont {Dai}\ \emph {et~al.}(2020)\citenamefont {Dai},
  \citenamefont {Agarwal},\ and\ \citenamefont {Cho}}]{dai2020nanoscale}%
  \BibitemOpen
  \bibfield  {author} {\bibinfo {author} {\bibfnamefont {C.}~\bibnamefont
  {Dai}}, \bibinfo {author} {\bibfnamefont {K.}~\bibnamefont {Agarwal}}, \ and\
  \bibinfo {author} {\bibfnamefont {J.-H.}\ \bibnamefont {Cho}},\ }\href@noop
  {} {\bibfield  {journal} {\bibinfo  {journal} {MRS Advances}\ }\textbf
  {\bibinfo {volume} {5}},\ \bibinfo {pages} {3507} (\bibinfo {year}
  {2020})}\BibitemShut {NoStop}%
\bibitem [{\citenamefont {Moscardi}\ \emph {et~al.}(2021)\citenamefont
  {Moscardi}, \citenamefont {Lanzani}, \citenamefont {Patern{\`o}},\ and\
  \citenamefont {Scotognella}}]{moscardi2021stimuli}%
  \BibitemOpen
  \bibfield  {author} {\bibinfo {author} {\bibfnamefont {L.}~\bibnamefont
  {Moscardi}}, \bibinfo {author} {\bibfnamefont {G.}~\bibnamefont {Lanzani}},
  \bibinfo {author} {\bibfnamefont {G.~M.}\ \bibnamefont {Patern{\`o}}}, \ and\
  \bibinfo {author} {\bibfnamefont {F.}~\bibnamefont {Scotognella}},\
  }\href@noop {} {\bibfield  {journal} {\bibinfo  {journal} {Applied Sciences}\
  }\textbf {\bibinfo {volume} {11}},\ \bibinfo {pages} {2119} (\bibinfo {year}
  {2021})}\BibitemShut {NoStop}%
\bibitem [{\citenamefont {Mandal}\ \emph {et~al.}(2023)\citenamefont {Mandal},
  \citenamefont {De}, \citenamefont {Lakshan}, \citenamefont {Sarfaraj},
  \citenamefont {Hazra}, \citenamefont {Dey},\ and\ \citenamefont
  {Mukhopadhyay}}]{RefA4}%
  \BibitemOpen
  \bibfield  {author} {\bibinfo {author} {\bibfnamefont {M.}~\bibnamefont
  {Mandal}}, \bibinfo {author} {\bibfnamefont {P.}~\bibnamefont {De}}, \bibinfo
  {author} {\bibfnamefont {S.}~\bibnamefont {Lakshan}}, \bibinfo {author}
  {\bibfnamefont {M.~N.}\ \bibnamefont {Sarfaraj}}, \bibinfo {author}
  {\bibfnamefont {S.}~\bibnamefont {Hazra}}, \bibinfo {author} {\bibfnamefont
  {A.}~\bibnamefont {Dey}}, \ and\ \bibinfo {author} {\bibfnamefont
  {S.}~\bibnamefont {Mukhopadhyay}},\ }\href@noop {} {\bibfield  {journal}
  {\bibinfo  {journal} {Journal of Optics}\ }\textbf {\bibinfo {volume} {2}},\
  \bibinfo {pages} {603} (\bibinfo {year} {2023})}\BibitemShut {NoStop}%
\bibitem [{\citenamefont {Afsari}\ and\ \citenamefont {Sarraf}(2020)}]{RefA3}%
  \BibitemOpen
  \bibfield  {author} {\bibinfo {author} {\bibfnamefont {A.}~\bibnamefont
  {Afsari}}\ and\ \bibinfo {author} {\bibfnamefont {M.~J.}\ \bibnamefont
  {Sarraf}},\ }\href@noop {} {\bibfield  {journal} {\bibinfo  {journal}
  {Superlattices and Microstructures}\ }\textbf {\bibinfo {volume} {138}},\
  \bibinfo {pages} {106362} (\bibinfo {year} {2020})}\BibitemShut {NoStop}%
\bibitem [{\citenamefont {Shang}\ \emph {et~al.}(2016)\citenamefont {Shang},
  \citenamefont {Zhang}, \citenamefont {Wang},\ and\ \citenamefont
  {Li}}]{RefJ4}%
  \BibitemOpen
  \bibfield  {author} {\bibinfo {author} {\bibfnamefont {S.}~\bibnamefont
  {Shang}}, \bibinfo {author} {\bibfnamefont {Q.}~\bibnamefont {Zhang}},
  \bibinfo {author} {\bibfnamefont {H.}~\bibnamefont {Wang}}, \ and\ \bibinfo
  {author} {\bibfnamefont {Y.}~\bibnamefont {Li}},\ }\href@noop {} {\bibfield
  {journal} {\bibinfo  {journal} {Journal of Colloid and Interface Science}\
  }\textbf {\bibinfo {volume} {483}},\ \bibinfo {pages} {11} (\bibinfo {year}
  {2016})}\BibitemShut {NoStop}%
\bibitem [{\citenamefont {Xu}\ \emph {et~al.}(2002)\citenamefont {Xu},
  \citenamefont {Friedman}, \citenamefont {Humfeld}, \citenamefont {Majetich},\
  and\ \citenamefont {Asher}}]{RefJ5}%
  \BibitemOpen
  \bibfield  {author} {\bibinfo {author} {\bibfnamefont {X.}~\bibnamefont
  {Xu}}, \bibinfo {author} {\bibfnamefont {G.}~\bibnamefont {Friedman}},
  \bibinfo {author} {\bibfnamefont {K.~D.}\ \bibnamefont {Humfeld}}, \bibinfo
  {author} {\bibfnamefont {S.~A.}\ \bibnamefont {Majetich}}, \ and\ \bibinfo
  {author} {\bibfnamefont {S.~A.}\ \bibnamefont {Asher}},\ }\href@noop {}
  {\bibfield  {journal} {\bibinfo  {journal} {Chemistry of Materials}\ }\textbf
  {\bibinfo {volume} {14}},\ \bibinfo {pages} {1249} (\bibinfo {year}
  {2002})}\BibitemShut {NoStop}%
\bibitem [{\citenamefont {Ge}\ and\ \citenamefont {Yin}(2008)}]{RefJ6}%
  \BibitemOpen
  \bibfield  {author} {\bibinfo {author} {\bibfnamefont {J.}~\bibnamefont
  {Ge}}\ and\ \bibinfo {author} {\bibfnamefont {Y.}~\bibnamefont {Yin}},\
  }\href@noop {} {\bibfield  {journal} {\bibinfo  {journal} {Chemistry of
  Materials}\ }\textbf {\bibinfo {volume} {20}},\ \bibinfo {pages} {3485}
  (\bibinfo {year} {2008})}\BibitemShut {NoStop}%
\bibitem [{\citenamefont {Ravi}\ and\ \citenamefont
  {Karthikeyan}(2013)}]{RefJ7}%
  \BibitemOpen
  \bibfield  {author} {\bibinfo {author} {\bibfnamefont {S.}~\bibnamefont
  {Ravi}}\ and\ \bibinfo {author} {\bibfnamefont {A.}~\bibnamefont
  {Karthikeyan}},\ }\href@noop {} {\bibfield  {journal} {\bibinfo  {journal}
  {Adv. Mat. Lett}\ }\textbf {\bibinfo {volume} {4}},\ \bibinfo {pages} {562}
  (\bibinfo {year} {2013})}\BibitemShut {NoStop}%
\bibitem [{\citenamefont {Xue}\ \emph {et~al.}(2017)\citenamefont {Xue},
  \citenamefont {Liu},\ and\ \citenamefont {Furlani}}]{xue2017theoretical}%
  \BibitemOpen
  \bibfield  {author} {\bibinfo {author} {\bibfnamefont {X.}~\bibnamefont
  {Xue}}, \bibinfo {author} {\bibfnamefont {K.}~\bibnamefont {Liu}}, \ and\
  \bibinfo {author} {\bibfnamefont {E.~P.}\ \bibnamefont {Furlani}},\
  }\href@noop {} {\bibfield  {journal} {\bibinfo  {journal} {The Journal of
  Physical Chemistry C}\ }\textbf {\bibinfo {volume} {121}},\ \bibinfo {pages}
  {9489} (\bibinfo {year} {2017})}\BibitemShut {NoStop}%
\bibitem [{\citenamefont {Hu}\ \emph {et~al.}(2015)\citenamefont {Hu},
  \citenamefont {He}, \citenamefont {Han}, \citenamefont {Wang},\ and\
  \citenamefont {Yin}}]{RefJ8}%
  \BibitemOpen
  \bibfield  {author} {\bibinfo {author} {\bibfnamefont {Y.}~\bibnamefont
  {Hu}}, \bibinfo {author} {\bibfnamefont {L.}~\bibnamefont {He}}, \bibinfo
  {author} {\bibfnamefont {X.}~\bibnamefont {Han}}, \bibinfo {author}
  {\bibfnamefont {M.}~\bibnamefont {Wang}}, \ and\ \bibinfo {author}
  {\bibfnamefont {Y.}~\bibnamefont {Yin}},\ }\href@noop {} {\bibfield
  {journal} {\bibinfo  {journal} {Nano Research}\ }\textbf {\bibinfo {volume}
  {8}},\ \bibinfo {pages} {611} (\bibinfo {year} {2015})}\BibitemShut {NoStop}%
\bibitem [{\citenamefont {Bohren}\ and\ \citenamefont {Huffman}(1998)}]{RefB2}%
  \BibitemOpen
  \bibfield  {author} {\bibinfo {author} {\bibfnamefont {C.~F.}\ \bibnamefont
  {Bohren}}\ and\ \bibinfo {author} {\bibfnamefont {D.~R.}\ \bibnamefont
  {Huffman}},\ }\href@noop {} {\emph {\bibinfo {title} {Absorption and
  Scattering of Light by Small Particles}}},\ \bibinfo {edition} {1st}\ ed.\
  (\bibinfo  {publisher} {WILEY-VCH Verlag GmbH \& Co},\ \bibinfo {year}
  {1998})\BibitemShut {NoStop}%
\bibitem [{\citenamefont {Mendoza-Herrera}\ \emph {et~al.}(2016)\citenamefont
  {Mendoza-Herrera}, \citenamefont {Scaffardi},\ and\ \citenamefont
  {Schinca}}]{RefJ9}%
  \BibitemOpen
  \bibfield  {author} {\bibinfo {author} {\bibfnamefont {L.~J.}\ \bibnamefont
  {Mendoza-Herrera}}, \bibinfo {author} {\bibfnamefont {L.~B.}\ \bibnamefont
  {Scaffardi}}, \ and\ \bibinfo {author} {\bibfnamefont {D.~C.}\ \bibnamefont
  {Schinca}},\ }\href@noop {} {\bibfield  {journal} {\bibinfo  {journal} {RSC
  Adv.}\ }\textbf {\bibinfo {volume} {6}},\ \bibinfo {pages} {110471} (\bibinfo
  {year} {2016})}\BibitemShut {NoStop}%
\bibitem [{\citenamefont {Xue}\ \emph {et~al.}(2025)\citenamefont {Xue},
  \citenamefont {Liu},\ and\ \citenamefont {Furlani}}]{RefJ18}%
  \BibitemOpen
  \bibfield  {author} {\bibinfo {author} {\bibfnamefont {X.}~\bibnamefont
  {Xue}}, \bibinfo {author} {\bibfnamefont {K.}~\bibnamefont {Liu}}, \ and\
  \bibinfo {author} {\bibfnamefont {E.~P.}\ \bibnamefont {Furlani}},\
  }\href@noop {} {\bibfield  {journal} {\bibinfo  {journal} {The Journal of
  Physical Chemistry C}\ } (\bibinfo {year} {2025})}\BibitemShut {NoStop}%
\end{thebibliography}%

\end{document}